\documentclass[conference,a4paper]{IEEEtran}

  	\usepackage[pdftex]{graphicx}
        \usepackage[tight,footnotesize]{subfigure}
  	\graphicspath{{../pdf/}{../jpeg/}}
	\DeclareGraphicsExtensions{.pdf,.jpeg,.png}
	\usepackage[cmex10]{amsmath}
	\usepackage{mathabx}
	\usepackage{algorithmic}
	\usepackage{array}
        \usepackage{tabularx}
	\usepackage{mdwmath}
	\usepackage{mdwtab}
	\usepackage{eqparbox}
	\usepackage{url}
        \usepackage{romannum}
        \usepackage{xcolor}
        \usepackage{balance}
        \usepackage{cite}
        \usepackage{enumitem}
        \usepackage{multirow}
        \usepackage[a4paper,left=1.7cm,right=1.6cm,top=2cm,bottom=2.6cm,headheight=0pt,footskip=0pt]{geometry}

\begin{document}

\title{Tuning of Ray-Based Channel Model for 5G Indoor Industrial Scenarios}
\author{Gurjot Singh Bhatia, Yoann Corre}

\author{\IEEEauthorblockN{
Gurjot Singh Bhatia\IEEEauthorrefmark{1}\IEEEauthorrefmark{2},   
Yoann Corre\IEEEauthorrefmark{1},   
M. Di Renzo\IEEEauthorrefmark{2}}                                     
\IEEEauthorblockA{\IEEEauthorrefmark{1} SIRADEL, Saint-Gregoire, France \\
\IEEEauthorrefmark{2} Universit\'e Paris-Saclay, CNRS, CentraleSup\'elec, Laboratoire des Signaux et Syst\`emes, Gif-sur-Yvette, France\\
gsbhatia@siradel.com}
}

\maketitle

\begin{abstract}
This paper presents an innovative method that can be used to produce deterministic channel models for 5G industrial internet-of-things (IIoT) scenarios. Ray-tracing (RT) channel emulation can capture many of the specific properties of a propagation scenario, which is incredibly beneficial when facing various industrial environments and deployment setups. But the environment's complexity, composed of many metallic objects of different sizes and shapes, pushes the RT tool to its limits. In particular, the scattering or diffusion phenomena can bring significant components. Thus, in this article, the Volcano RT channel simulation is tuned and benchmarked against field measurements found in the literature at two frequencies relevant to 5G industrial networks: 3.7 GHz (mid-band) and 28 GHz (millimeter-wave (mmWave) band), to produce calibrated ray-based channel model. Both specular and diffuse scattering contributions are calculated. Finally, the tuned RT data is compared to measured large-scale parameters, such as the power delay profile (PDP), the cumulative distribution function (CDF) of delay spreads (DSs), both in line-of-sight (LoS) and non-LoS (NLoS) situations and relevant IIoT channel properties are further explored.

\end{abstract}


\vskip0.5\baselineskip

\IEEEpeerreviewmaketitle


\section{Introduction}
The 5G mobile communication was developed to enhance mobile networks' broadband capabilities and supply improved wireless access to a wide range of industry verticals, such as the manufacturing, automotive, and agricultural sectors \cite{5g2020key}. Industrial environments are considered severe from the point of view of electromagnetic (EM) wave propagation. A large number of obstructions and fading fluctuations may cause degraded signal and system reliability. Hence, radio channel characterization is critical for designing radio communication systems for future smart factories. 


A channel model is known as an abstract and simplified approach to mathematically or computationally reproduce the main characteristics of an actual channel and evaluate the impact on the performance of a specific wireless technology.

Empirical channel models rely on wide-band channel measurements to characterize propagation by statistically assessing wide-band channel properties and then formulating mathematical relationships and equations to derive important variables like path loss, DS, angular spread, and so forth. Some popular empirical path loss models for indoor scenarios are: Log-Normal (Large-scale) Shadowing model, Alpha Beta Gamma (ABG) model, and Close-in (CI) free space path loss model. The 3rd Generation Partnership Project (3GPP) in the technical report (TR) 38.901 version 16.1.0 \cite{3gpp2019study} also supports the ABG model and CI free space path loss model for indoor scenarios, such as Indoor Hotspot - Office (InH) and Indoor Factory (InF) scenarios.

Non-geometric Stochastic Channel Models (NGSCMs) illustrate and determine physical parameters entirely stochastically by dictating underlying probability distribution functions without assuming an underlying geometry (examples are the Saleh-Valenzuela or the Zwick model). Whereas, Geometry-based Stochastic Channel Models (GSCMs) rely on some geometrical assumptions, and the propagation parameters are at least partially stochastic and specified by probability distributions \cite{zwick2002stochastic}. The COST models, the 3GPP Spatial Channel Model (SCM), and Wireless World Initiative New Radio (WINNER) are some examples of GSCMs.

Stochastic channel models (SCMs) do not aim to reproduce channel responses at a particular site but are plausible and realistic given an imaginary environment. SCMs leverage radio measurements such as wide-band channel measurements to derive statistical distributions for large-scale parameters and distribution laws for multi-path components (MPCs).

Deterministic channel modeling is another approach to simulate radio propagation based on Maxwell's equations. RT is a commonly-used technique in deterministic channel modeling. Its algorithm starts by finding all the possible geometrical rays between a transmitter and a receiver for a given number of allowed interactions. Then, the calculation of the rays' (EM field) contributions is based on the geometrical optics (GO), uniform theory of diffraction for diffraction (UTD), and effective roughness theory (ER), assuming that the far-field conditions are met. The RT can provide path loss data, angle-of-arrival (AoA), angle-of-departure (AoD), time delay, optical visibility (LoS or NLoS), etc.

RT, similar to the finite element method, can require a lot of computational resources to produce precise results, especially for complex problems. This is particularly consequential in industrial environments with many complicated structures and objects with complex geometries. As a result, RT simulations can become computationally intensive for InF scenarios.

Recently, 3GPP in the TR 38.901 version 16.1.0 \cite{3gpp2019study}, has proposed models for typical smart factory environments called InF and IIoT channel models. These channel models rely significantly on measurement data collected in typical propagation conditions. Several measurement campaigns in InF scenarios can be found in the literature. In \cite{8377337}, the authors try to assess the channel propagation at 28 and 60 GHz frequencies for light and heavy industrial layouts. In \cite{9815573}, the authors demonstrate massive multi-input multi-output (mMIMO) channel characteristics based on channel sounding measurements carried out in an industrial environment at frequency ranges from 26 to 30 GHz. In \cite{schmieder2019directional}, the authors perform directional wide-band channel measurements at 28 GHz in an industrial environment. The work in \cite{schmieder2020measurement} explains a wide-band channel measurement campaign at 3.7 and 28 GHz with direction-of-arrival information at 28 GHz. Such campaigns typically use empirical models and analyze channel parameters such as path loss, PDP, AoA, AoD, LoS probability, and Root Mean Square (RMS) DS, among other relevant parameters.

Stochastic and empirical channel models rely heavily on wide-band field measurements \cite{sun2016propagation,schmieder2019directional, schmieder2020measurement}. Field measurements, such as channel-sounding measurements, are used to get precise data about the radio channel. They pose several difficulties in complexity, cost, and representability, which need efficient radio propagation models to be developed as workable replacements. Another challenge for those models is that propagation in InF scenarios is more site-specific than in usual residential or office environments \cite{8377337, schmieder2020measurement}. For instance, in InF scenarios, metallic machines are one of the most common objects in the environment. The huge bodies of such machines can become major blockers in the NLoS case. The smooth metallic surface creates many reflections, and the large body prevents the signal from propagating directly. The InF environment with a large number of machines complicates the radio propagation in the factory.

Given the drawback of field measurements, the site-specific nature of InF scenarios, and the recent interest in mmWave frequency bands for wireless networks, many channel and network performance metrics for mmWave communications have been lately generated through RT \cite{gougeon2019ray, charbonnier2020calibration}. Hence, ray-based channel models can be considered an interesting radio propagation and characterization approach in InF scenarios. 

In this paper, we present an approach that can be used to produce a site-specific calibrated ray-based channel model for a typical 5G InF scenario. This paper aims to reduce the dependency on complicated field measurements. This will allow a tool to create site-specific channel data in factory scenarios and a database of channel samples that may be used for research studies. There have been earlier attempts to calibrate ray-based models for urban scenarios using measurement data like in \cite{charbonnier2020calibration}. Still, to the best of our knowledge, there has not been any such attempt for industrial environments yet. 

This paper is structured as follows. Section II briefly describes the various benefits of RT for InF scenarios. Section III describes the approach used for the validation of RT against measurements. Section IV portrays some ideas for exploiting the calibrated RT tool. Finally, Section V presents the conclusion and future perspective of this work.


\section{Benefits of Ray-tracing for InF scenarios}

The RT approach presents some benefits to realize the next-generation channel models for future smart factories. These benefits can provide a new paradigm to study the signal propagation characteristics and model the wireless channel in industrial scenarios. 
\begin{itemize}[leftmargin=0.5cm]
   \item RT is a powerful technique for predicting ``specular MPCs" (SMPCs: specular reflections and diffractions) and ``dense MPCs" (DMPCs: diffuse scattering) in complex industrial settings. RT enables deterministic modeling of ray paths, making it valuable for channel modeling, network planning, and optimization, particularly at mmWave frequencies where there are a limited number of such paths.
   \item RT requires detailed modeling of the propagation environment, incorporating precise dimensions and dielectric properties of objects. Hence, RT models can be tailored to specific scenarios, effectively capturing the site-specific characteristics of indoor industrial environments. 
   \item The RT model can be calibrated \cite{charbonnier2020calibration}. Once a RT model is calibrated and validated for a specific scenario, it can predict channel parameters for different positions of the base station (BS) and the user terminal (UT), and predict coverage maps and correlation properties. It can also predict MIMO channel properties, spectral efficiency, and data rates.
   \item The validated RT model then can provide received power and other large-scale parameters at arbitrary locations for radio coverage planning and network optimization of a communication system.
   \item Spatial and time variability of the channel in terms of the movement of the transmitter, receiver, and other objects (such as mobile robots, automated guided vehicles (AGVs), drones, and humans.) in the environment can also be included.
   \item It is relatively easy to include beyond 5G enabling technologies such as reconfigurable intelligent surfaces (RIS) into RT models. For instance, the effective roughness theory used to model diffuse scattering interactions can be used to model the anomalous reflections from RIS \cite{degli2022reradiation}.
 \end{itemize}

\section{Ray-tracing validated against measurements}

This section will focus on the benchmarking attempt against the measurement results found in the literature. In \cite{schmieder2020measurement}, the authors present a wide-band channel measurement campaign at 3.7 and 28 GHz with direction-of-arrival information at 28 GHz. Large-scale channel parameters are evaluated using empirical channel models, and the results are compared to the 3GPP TR 38.901 InF model. It is an example of an industrial measurement campaign and a typical use case for InF scenarios. Hence, it will be a reference to benchmark and tune our ray-based model.

\begin{figure}[ht!] 
\centering
\includegraphics[width=3.35in]{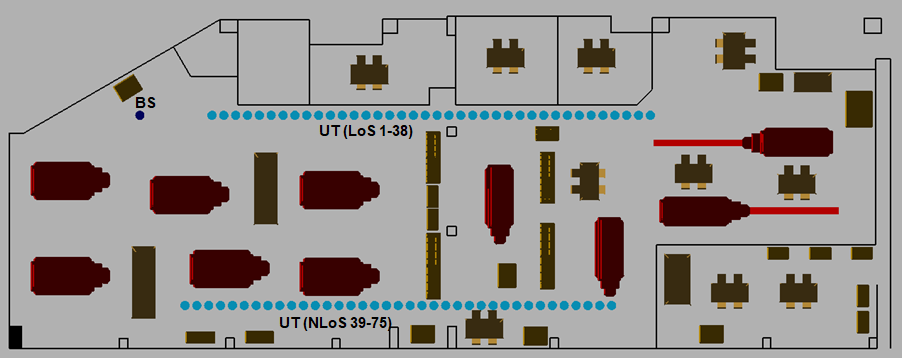}
\caption{3D digital model of the measurement site in channel emulator.}
\label{2}
\end{figure}

The 3D digital model of the measurement scenario, as shown in Fig. \ref{2}, was created just from the 2D floor plan available in the literature. We did not have access to a detailed description, photos of the measurement site, or exact specifications of the measurement setup. Thus, the chosen objects, their material, and their dimensions are a guess based on the reference paper and analysis of typical objects found in such factories. The exact location of different receiver positions was also unknown. 


\subsection{The 3D digital scenario and simulation settings}

The BS antenna used in simulations is a single element, $\frac{\lambda}{2}$ vertically polarized omni-directional antenna. The transmitted power was kept at 0 dBm to analyze only the propagation channel's impact. The UT antenna is also a single element, $\frac{\lambda}{2}$ vertically polarized omni-directional antenna. The BS is deployed at a fixed position at a height of 1.85 m. The UTs are deployed at 75 different positions: 1-38 LoS (step size = 1 m) and 39-75 NLoS (step size = 1 m), as shown in Fig. \ref{2}. All the UTs are at a height of 1.44 m. The digital model is 74.4 m long, 24.4 m wide, and 4.6 m high. Table \ref{table:2} gives the specifications of the 3D digital scenario, while table \ref{table:1} gives an overview of the various objects in the 3D digital scenario.

\begin{table}[ht!]
 \centering
 \caption{Specifications of the 3D digital scenario.}
 \label{table:2}
 \begin{tabular}{|>{\centering\arraybackslash\vspace{0.5ex}}m{3.5cm}<{\vspace{0.5ex}}|>{\centering\arraybackslash\vspace{0.5ex}}m{3.5cm}<{\vspace{0.5ex}}|} 
  \hline
  \textbf{Parameters} & \textbf{Values} \\
  \hline
  Hall size ($m$) & $74.4 \times 24.4$ \\ 
  \hline
  Hall height ($m$) & $4.6$ \\ 
  \hline
  Clutter density & Low clutter density $\approx$ $18.33\%$ \\ 
  \hline
  Average clutter height ($m$) & 2 \\ 
  \hline
  Carrier frequency: bandwidth & $3.7$ GHz: $80$ MHz, $28$ GHz: $100$ MHz \\ 
  \hline
 \end{tabular}
\end{table}

 \begin{table}[ht!]
 \centering
 \vspace{0.2cm}
 \caption{Various objects in the 3D digital scenario.}
 \label{table:1}
 \begin{tabular}{|>{\centering\arraybackslash\vspace{0.5ex}}m{3cm}<{\vspace{0.5ex}}|>{\centering\arraybackslash\vspace{0.5ex}}m{3cm}<{\vspace{0.5ex}}|} 
  \hline
  \textbf{Object} & \textbf{Dimension (\textbf{$m$})}  \\ 
  \hline
  Machines & $6.7 \times 2.4 \times 2$ \\  
  \hline
  Storage racks & $2.3 \times 1.1 \times 4$ \\  
  \hline
  Cupboards & $1.8 \times 1 \times 3.6$ \\ 
  \hline
  Industrial worktables & $6 \times 2 \times 0.8$ \\ 
  \hline
  Tables and Chairs & $3.2 \times 1.6 \times 0.5$ \\ 
  \hline
  Crates & $2 \times 1.5 \times 1$ \\ 
  \hline
 \end{tabular}
 \end{table}



After creating a digital model of the measurement site, point-to-multi-points (P2MP) simulations were realized for positions 1-38 for the LoS case and 39-75 for the NLoS case, using Volcano Flex \cite{Siradel} RT. Volcano Flex is a time-efficient propagation model based on the ray-launching (also known as the shooting-and-bouncing) approach, capable of predicting deterministic path-loss in any small-scale urban or indoor scenario. It can provide channel properties and 3D multi-path trajectories. By default, two reflections and one diffraction are allowed for each ray.

For P2MP simulation results, various channel parameters, such as received power, PDP, channel impulse response (CIR), transfer function (TF), Azimuth AoA, Elevation AoD, and many more, can be calculated. After the initial analysis, the simulation results of position 1 (LoS) and 39 (NLoS) were compared with the measured results. The multi-paths richness was observed to be strongly underestimated, and low-power components at higher delays were missing. Hence, the number of allowed interactions was changed. The P2MP analysis was repeated with a maximum of three reflections, one diffraction allowed for each ray, and diffuse scattering from walls and machines were activated \cite{charbonnier2020calibration}. 

The PDPs showed significant improvement for most of the positions when the number of interactions was increased, but this increased the computation times as well. A point-to-point (P2P) link simulation with three reflections, one diffraction, and diffuse scattering from walls and machines took around two and a half hours compared to about thirty minutes if diffuse scattering was disabled. This time further decreases to a few seconds for the case of two reflections and one diffraction for each ray, but the results worsen significantly. Hence, the results could further be improved by increasing the number of interactions, but this would significantly increase the computation times.



 \subsection{Calibration of channel simulation results}
 
The calibration process aims to adjust the contribution of different types of ray interactions and minimize the difference between the simulated and measured results. 

As mentioned earlier, metallic machines are a common object in InF situations. Their enormous size makes them a major NLoS obstacle and scatterer for low-elevation user devices. Machines are not represented with all details but by a large simple block. They are associated with the material properties of a typical metal \cite{ITU}.

After visual inspection of predicted rays and initial channel parameters, it could be seen that the SMPCs (specular reflections and diffractions) from the machines were over-estimated. Hence, to decrease the impact of SMPCs just from the machines, the default metal properties were swapped to $\varepsilon_\mathrm{r}^{'} = 3$, $\varepsilon_\mathrm{r}^{''} = 0.1$ at 3.7 GHz and $\varepsilon_\mathrm{r}^{'} = 3$, $\varepsilon_\mathrm{r}^{''} = 0.09$ at 28 GHz, with a thickness of 40 cm. This would decrease the weight of the specular reflections and diffractions, and allow for some (minor) transmission, while the strength of the diffuse scattering remains unchanged. These equivalent material properties offer significant improvement. However, we did not exhaustively analyze, and further optimization remains. 

Then we used the same principle as in \cite{charbonnier2020calibration} with ray-path classification and calibration to better fit the measurements. PDPs (maximum power and power decay trend), DSs, and angular distributions are analyzed for this study.

\begin{figure}[ht!] 
\centering
\includegraphics[width=3.3in]{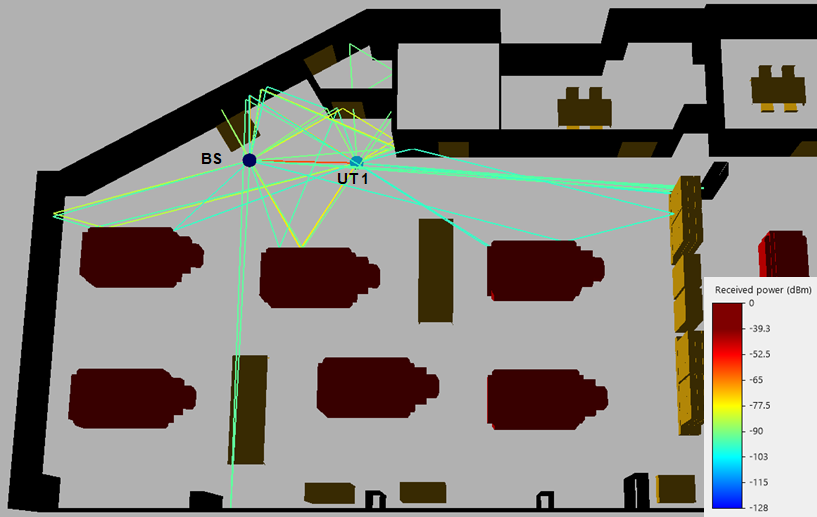}
\caption{Some of the predicted specular ray components for UT 1 (LoS) after initial calibration.}
\label{6}
\end{figure}

Furthermore, the diffraction components were decreased by a diffraction offset of 10 dB. The contribution of the DMPCs (diffuse scattering) from the channel simulations was weaker than the measured results. Hence, the DMPCs were increased by a diffuse scattering offset of 12 dB. This coarse correction has no apparent physical justification but shows promising results and will permit the realization of realistic studies.

\subsection{Comparison of calibrated and measured results} 

\subsubsection{Power Delay Profile}

\begin{figure}[ht!] 
\centering
\includegraphics[width=3.4in]{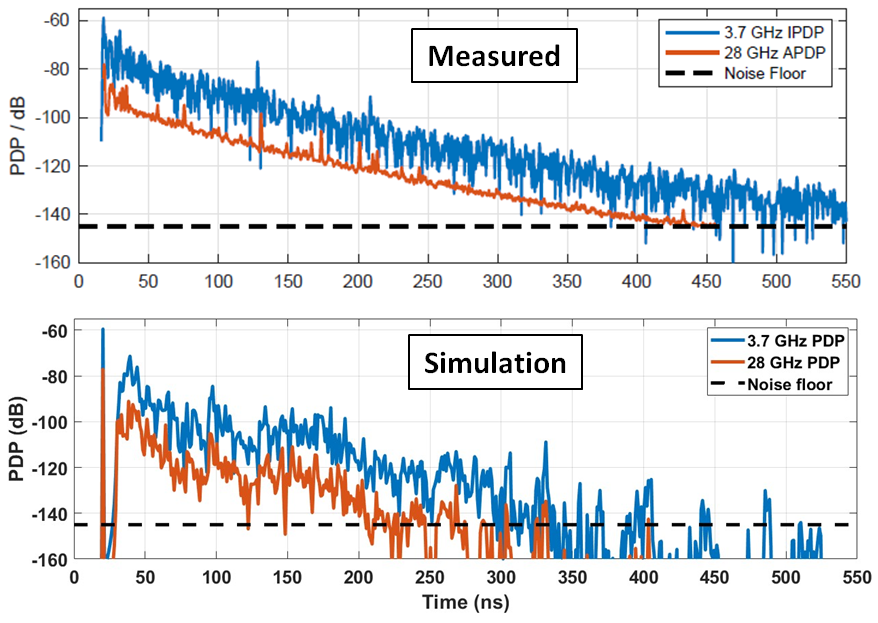}
\caption{Comparison between the measured PDP (top picture, taken from \cite{schmieder2020measurement}) and the simulated PDP (bottom), for LoS case.}
\label{fig:PDP_comp_LOS}
\end{figure}

Fig. \ref{fig:PDP_comp_LOS} (top) \cite{schmieder2020measurement} shows the instantaneous PDP (IPDP) and averaged PDP (APDP) from the measurement data at 3.7 and 28 GHz, respectively, for the LoS case. The LoS PDP corresponds to a measurement point with a BS-UT distance of 5.2 m, resulting in a time of flight (ToF) delay of 17.5 ns. At 3.7 GHz, the measured LoS component was received with a power of -58.8 dBm, and at 28 GHz, with a power of -78.1 dBm. Fig. \ref{fig:PDP_comp_LOS} (bottom) also shows the simulated LoS PDP from the RT model. It corresponds to a BS-UT distance of 6.1 m, resulting in a ToF delay of 20.3 ns. This component was received with a power of -59.6 dBm, and at 28 GHz, with a power of -77.1 dBm. In LoS case for both measured and simulated results, several SMPCs can be seen together with DMPCs for both frequencies. For measurement results, they reach the noise floor of -145 dBm at a delay of about 450 ns at 28 GHz and 550 ns at 3.7 GHz. 

Thanks to calibration, the average trend of the simulated PDP is similar to the measurement until 180 ns. Beyond this value, the simulated power decrease accelerates; barely any propagation path exceeds the noise floor at delays greater than 320 ns and 400 ns at 28 GHz and 3.7 GHz, respectively. 


\begin{figure}[ht!] 
\centering
\includegraphics[width=3.4in]{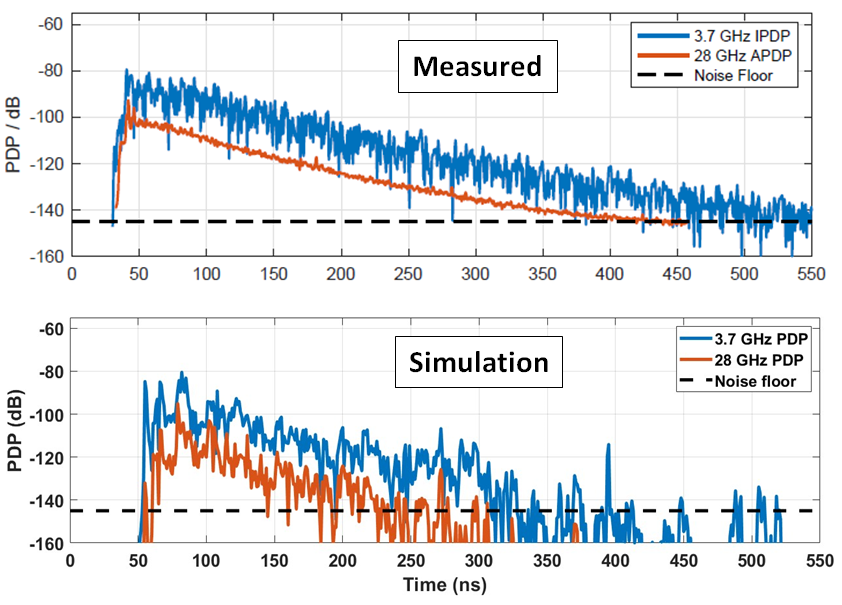}
\caption{Comparison between the measured PDP (top picture, taken from \cite{schmieder2020measurement}) and the simulated PDP (bottom), for NLoS case.}
\label{fig:PDP_comp_NLOS}
\end{figure}

Fig. \ref{fig:PDP_comp_NLOS} (top) \cite{schmieder2020measurement} shows the IPDP and APDP from the measurement data at 3.7 and 28 GHz, respectively, for NLoS case. The NLoS PDP corresponds to a measurement point with a BS-UT distance of 9.9 m, resulting in a time of flight delay of 33 ns. Fig. \ref{fig:PDP_comp_NLOS} (bottom) also shows the simulated LoS PDP. It corresponds to a BS-UT distance of 16.5 m, resulting in a ToF delay of 54.9 ns. For measurement results, only a few strong SMPCs can be seen at 3.7 GHz with a delay of 40 ns and a power of -79.4 dBm. It is difficult to distinguish between specular and diffuse components, even in a short delay range. Besides, in the simulation, we observe some dominant specular ray-paths at 65 and 80 ns with a power close to -80 dBm. The maximum predicted power is consistent with the measurement. However, the density of components detected at this level is underestimated. The reason may come from the simplification of the machinery representation.

As shown in Fig. \ref{fig:PDP_comp_LOS} and \ref{fig:PDP_comp_NLOS}, the LoS and NLoS PDPs from measurement and simulated results follow the same trend with some disagreements that can also be attributed to the lack of detailed information about the measurement setup and scenario. Ray-based channel models, especially for the industrial environment, are highly scenario specific. Hence, the lack of proper details about various industrial objects' size, shape, position, and other aspects can change RT results, especially for NLoS cases and mmWave propagation, where DMPCs play a significant role.
\vspace{0.1in}

\subsubsection{RMS DS and Angular Spread}

\begin{figure}[ht!] 
\centering
\includegraphics[width=3.3in]{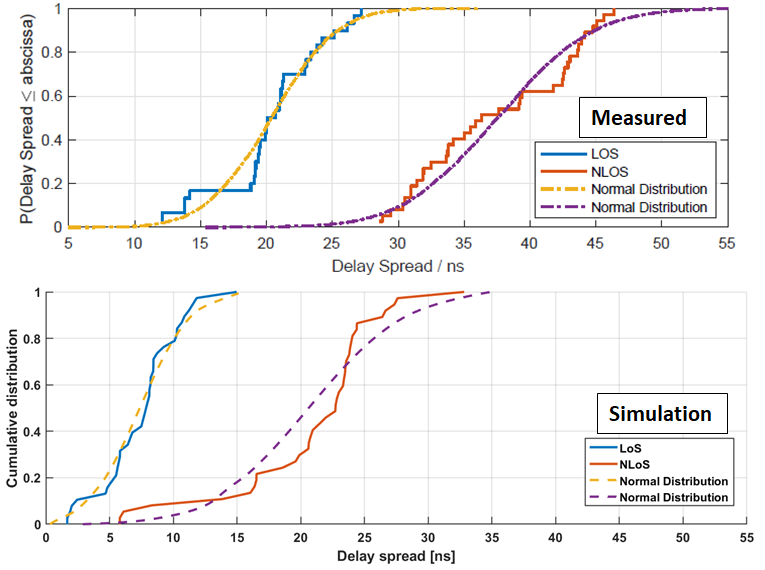}
\caption{Comparison between the measured CDF of RMS DS (top picture, taken from \cite{schmieder2020measurement}) and the simulated CDF of RMS DS (bottom),  for LoS and NLoS case at 3.7 GHz.}
\label{fig:PDP_CDF}
\end{figure}

\begin{figure}[ht!] 
\centering
\includegraphics[width=3.35in]{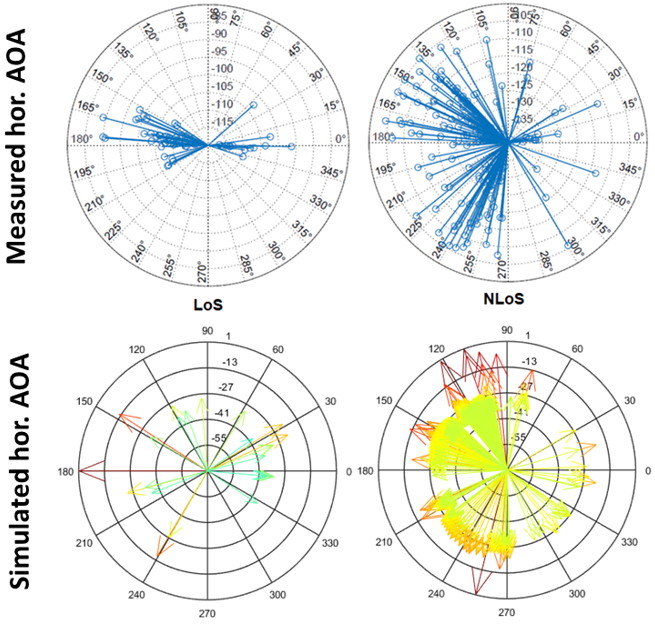}
\caption{Comparison between the measured (top picture, taken from \cite{schmieder2020measurement}) and the simulated horizontal AoA power profile at 28 GHz.}
\label{fig:PDP_AoA}
\end{figure}

As shown in Fig. \ref{fig:PDP_CDF}, the CDF of DS was calculated using the calibrated RT results and compared with the measurement results at 3.7 GHz. For LoS case, the DS for the simulated scenario is smaller than the measured DS, due to underestimated power at higher delays. For the NLoS case, the observed difference is the opposite; this can be partly attributed to the fact the BS-UT distances in the digital scenario are almost 1.5 times larger than in the actual measurement scenario. The comparison also reinstates that it is hard to distinguish between specular and diffuse components, even in a short delay range. This complicates choosing the optimal offset for different interactions and can undermine the multi-path richness of the channel. The horizontal (azimuth) AoA power spectrum was also calculated and compared to the measurements, as shown in Fig. \ref{fig:PDP_AoA}.

\section{Exploitation}

After its calibration, the RT tool can be exploited for various applications. We give some examples here.

\begin{figure}[ht!] 
\centering
\includegraphics[width=3.35in]{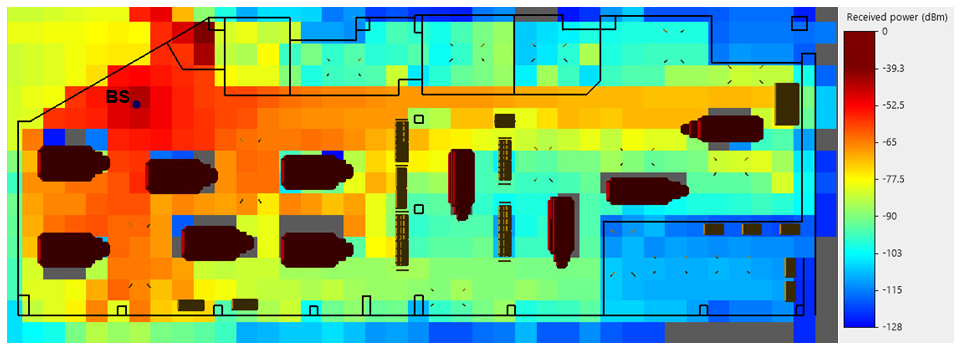}
\caption{Coverage map of the factory floor at 3.7 GHz.}
\label{13}
\end{figure}

Fig. \ref{13} shows the coverage map of the factory floor at 3.7 GHz with a resolution of 2 m. The coverage analysis can be performed with a finer resolution, but this would cost higher computation times. Such a map is interesting to complement the channel analysis; it provides a more extensive and comprehensible evaluation of the factory's power distribution beyond the scope of what can be observed solely through measurements. For instance, we can see different NLoS situations with higher or lower power degradation.

\begin{figure}[ht!] 
\centering
\includegraphics[width=3.4in]{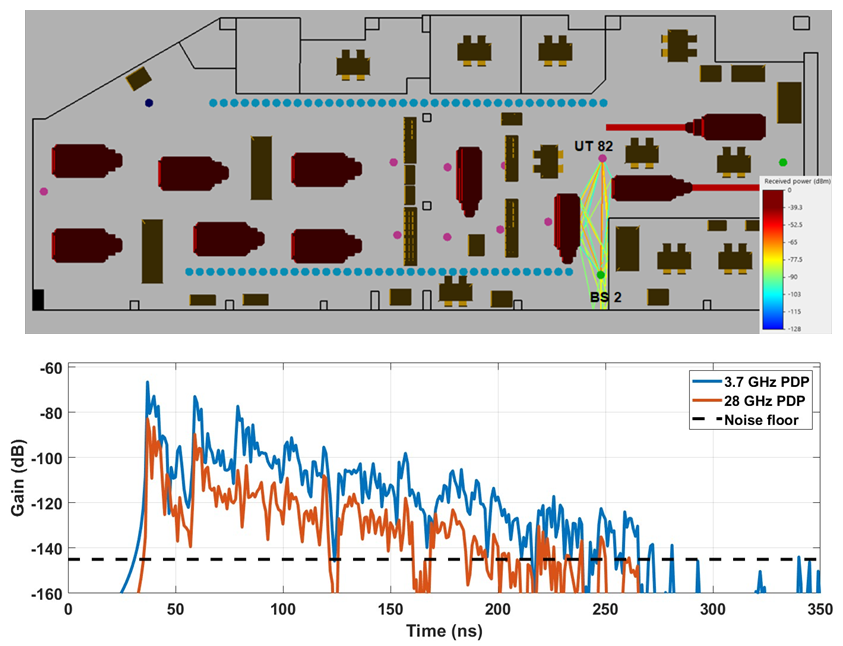}
\caption{Simulation results corresponding to the new BS 2 and UT 82 at 3.7 GHz.}
\label{fig:PDP_CDF_newpos}
\end{figure}

The calibrated model can also easily analyze channel properties between new BS-UT positions, as shown in Fig. \ref{fig:PDP_CDF_newpos}. This makes the ray-based model a flexible tool to complement the measurement-based channel characterization.

\begin{figure}[ht!] 
\centering
\includegraphics[width=3.45in]{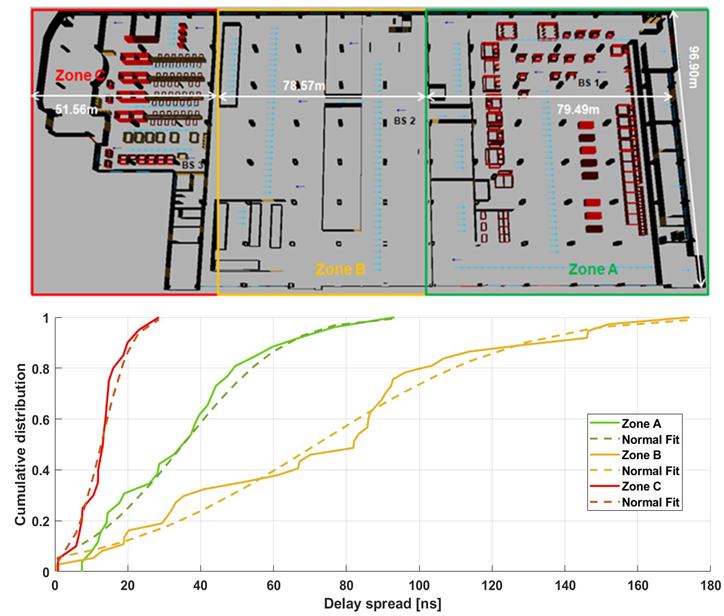}
\caption{A generic industrial scenario with CDFs of DS at 3.7 GHz of the three different zones for NLoS case.}
\label{fig:new_factory}
\end{figure}

Finally, Fig. \ref{fig:new_factory} (top) shows a factory scenario that the authors in \cite{eucnc_gurjot} use to investigate various channel emulation use cases, like simplifying the propagation model to improve the time efficiency. Based on a real factory setup, three distinct zones are identified. These zones differ from each other in terms of shape, size, object material, and clutter density. Zone A is the biggest. It is 79.5 m long and is characterized by machines and storage containers. Zone B is 78.6 m long and is empty primarily to facilitate the movement of factory workers and their loads, with two big lobbies connecting a few rooms. Zone C is 51.6 m long, with metallic lockers, wooden benches, and metallic housing units.

This scenario requires a separate in-depth study and analysis, but the first simulations conducted with the approach mentioned in this paper are reported here. Fig. \ref{fig:new_factory} (bottom) shows that the CDFs of DS vary in a significant manner from one zone to another. These results re-emphasize that the IIoT channel models are site-specific, and any change in the factory scenario can strongly impact the channel properties. Such a problem involving producing channel samples for InF scenarios can be easier to address using a RT tool. The approach suggested in this paper can be used to reduce the dependency on complicated channel measurements and provide more flexibility to produce channel models for InF and IIoT scenarios.


\section{Conclusions and future perspective}

This work tested and calibrated a ray-based channel model in an industrial scenario. The calibration steps helped improve the average trend of simulated PDPs, DSs, and angular distribution. Several aspects need further analysis, such as optimized material properties and extent of simplification of representation of various objects in the environment, more ways to increase the multi-path richness of the channel, and optimized diffraction and diffuse scattering offset. This work is undoubtedly one of the first studies directed toward producing a site-specific calibrated ray-based channel model for typical 5G InF scenarios.

It is worth noting that the calibrated ray-based channel model can be used to predict several other channel and network parameters for the given factory scenario, such as coverage maps, correlation properties, MIMO channel properties, spectral efficiency, signal-to-noise ratio, data rates, and so forth. Besides, it is relatively easy to include beyond 5G enabling technologies such as mMIMO and reconfigurable intelligent surfaces (RIS) into RT models. Considering the challenges presented by InF and IIoT scenarios, the ray-based model offers an intriguing supplementary or alternative option to empirical and stochastic methodologies.

The produced channel samples will be completed with additional scenarios and made publicly available with free access soon. 


\section*{Acknowledgment}

This work is part of a project that has received funding from the European Union’s Horizon 2020 research and innovation programme under the Marie Skłodowska Curie grant agreement No. 956670. We thank the Fraunhofer Heinrich Hertz Institute's support in understanding the measurement setup.



\end{document}